\documentstyle[prd,aps,floats]{revtex}
\begin{document}
\draft
%
%
\input epsf
\renewcommand{\topfraction}{0.8}
\twocolumn[\hsize\textwidth\columnwidth\hsize\csname
@twocolumnfalse\endcsname
\preprint{CERN-TH/96-358, SU-ITP-96-60, astro-ph/9612141}
\title{Tilted Hybrid Inflation}
\author{Juan Garc\'{\i}a-Bellido}
\address{Theory Division, CERN, CH-1211 Geneva 23, Switzerland}
\author{Andrei Linde}
\address{Physics Department, Stanford University,
Stanford CA 94305-4060, USA}

\date{December 13, 1996}
\maketitle
\begin{abstract}
  We propose a new version of the hybrid inflation scenario that
  produces a significantly `tilted' $n>1$ spectrum of curvature
  perturbations. This may happen in supersymmetric models where the
  inflaton field acquires a mass proportional to the Hubble constant,
  and in the models where this field non-minimally couples to gravity.
  A large tilt of the spectrum in this scenario is often accompanied
  by a considerable contribution of tensor perturbations to the
  temperature anisotropies. We also show that under certain conditions
  the spectrum of density perturbations in this scenario may have a
  minimum on an intermediate length scale.
\end{abstract}
\pacs{PACS numbers: 98.80.Cq \hspace{5mm}
  Preprint CERN-TH/96-358,~~SU-ITP-96-60 ~~~~~~astro-ph/9612141}

\vskip2pc]

The inflationary paradigm~\cite{book} not only provides a solution to
the classical problems of the hot big bang cosmology, but also
predicts an almost scale invariant spectrum of curvature perturbations
which could be responsible for the observed anisotropy of the cosmic
microwave background (CMB), as well as the origin of the large scale
structure. Until recently, observations of the temperature
fluctuations in the CMB provided just a few constraints on the
parameters of inflationary models, mainly from the amplitude and the
tilt of both scalar (density) and tensor (gravitational waves)
perturbations spectrum. Nowadays, with several different experiments
looking at different angular scales, we have much more information
about both spectra as well as other cosmological parameters like
$\Omega_0$, $H_0$, $\Omega_{\rm B}$, etc. In the near future, most of
these parameters will be known with better than a few percent
accuracy, thanks to the new generation of microwave anisotropy
satellites, {\it MAP}~\cite{MAP} and {\it COBRAS/SAMBA}~\cite{COBRAS}.
It is therefore the responsibility of theoretical cosmologists to
provide a variety of inflationary models with definite predictions,
which could be compared with the observational data.

In this paper, we consider a new model of hybrid
inflation~\cite{hybrid} with a large tilt for the spectral index of
density perturbations. Models of hybrid inflation are known to provide
a blue tilted spectrum of scalar perturbations, with negligible
contribution of gravitational waves~\cite{hybrid,LL93,CLLSW,GBW}.
However, in most of the hybrid inflation models developed so far the
tilt is extremely small; in order to achieve a considerable tilt of
the spectrum in hybrid inflation one needs to fine tune the parameters
of the model~\cite{GBW,Lyth}. In what follows we will introduce a new
class of hybrid inflation models where a significant tilt can be
achieved in a natural way.

Such a tilted model could be a simple way of implementing an open
inflation model~\cite{open,LM} together with a tilted spectrum, which
is one of the preferred choices suggested by present observations of
the CMB and large scale structure, see Ref.~\cite{WS}. We will deal
with such a possibility in a following publication~\cite{GBL}.

The simplest realization of hybrid inflation in the context of the
chaotic inflation scenario is provided by the potential~\cite{hybrid}
\begin{equation}\label{hybrid}
V(\phi,\psi) = {1\over4\lambda}\left(M^2-\lambda\psi^2 \right)^2
 + {1\over2}m^2\phi^2 + {1\over2}g^2\phi^2\psi^2 \, .
\end{equation}
The bare masses $m$ and $M$ of the scalar fields $\phi$ and $\psi$ are
``dressed'' by their mutual interaction. At large values of the
fields, their effective masses squared are both positive and the
potential has the symmetry $\psi \leftrightarrow -\psi$. At small
values of the field $\phi$, the potential has a maximum at
$\phi=\psi=0$ and a global minimum at $\phi=0, \psi=\psi_0\equiv
M/\sqrt\lambda$, where the above symmetry is broken.

Equations of motion for the homogeneous fields at the stage of their
slow rolling during inflation are
\begin{eqnarray}\label{EQM}
3H\dot\phi &=& - (m^2 +g^2\psi^2) \phi\, , \\
3H\dot\psi &=& (M^2 - g^2\phi^2 - \lambda\psi^2) \psi \, ,
\end{eqnarray}
where the Hubble constant at that stage is given by $H^2 = {8\pi V/
  3M^2_{\rm P}}$. Motion starts at large $\phi$, where the effective
mass squared of the $\psi$ field is positive and the field is sitting
at the minimum of the potential at $\psi=0$. As the field $\phi$
decreases, its quantum fluctuations produce an almost scale invariant
but slightly tilted spectrum of density
perturbations~\cite{hybrid,LL93,CLLSW,GBW,Lyth}.

During the slow-roll of the field $\phi$, the effective mass of the
triggering field is $m^2_\psi = g^2\phi^2 - M^2$. When the field
$\phi$ acquires the critical value $\phi_c \equiv M/g$, fluctuations
of the massless $\psi$ field trigger the symmetry breaking phase
transition that ends inflation. If the bare mass $M$ of the $\sigma$
field is large compared with the rate of expansion $H$ of the
universe, the transition will be instantaneous and inflation will end
abruptly~\cite{hybrid}. If on the contrary the bare mass $M$ is of
the order of $H$, then the transition will be very slow and there is
the possibility of having yet a few more $e$-folds of inflation after
the phase transition \cite{Guth,GBLW}.

At $\psi=0$ the inflaton potential is $V(\phi) = M^4/4\lambda +
m^2\phi^2/2$. Since the scalar field $\phi$ takes relatively small
values, for $m^2\ll g^2M^2/\lambda$ the energy density is dominated by
the vacuum energy,
\begin{equation}\label{H0}
H_0^2 =  {2\pi M^4\over3\lambda M^2_{\rm P}}\,,
\end{equation}
to very good accuracy. It is then possible to
integrate the evolution equation for $\phi$,
\begin{eqnarray}\label{soln}
\phi(N) &=& \phi_c\,\exp(rN)\, , \\
r &=& {3\over2} - \sqrt{{9\over4}-{m^2\over H_0^2}} \,
\simeq {m^2\over3H_0^2}\,,
\end{eqnarray}
where $N=H_0(t_e-t)$ is the number of $e$-folds to the end of
inflation at $\phi=\phi_c$, and we have used the approximation
$m\lesssim H_0$.

An important feature of the perturbation spectrum of the hybrid
inflation model is its growth at large $k$; for example, at the end of
inflation in this theory one has \cite{hybrid,LL93}:
\begin{equation}\label{denspert}
\frac{\delta\rho}{\rho} (k) =  {2\sqrt{6\pi} g M^5\over 5\lambda
\sqrt{\lambda} m^2 M_{\rm P}^3}\,\Big({k\over H_0}\Big)^{m^2\over
3H_0^2}\ ,
\end{equation}
which corresponds to a spectral index
\begin{equation}\label{index}
n \simeq 1 + {2m^2\over 3H_0^2} = 1 + {\lambda m^2
M_{\rm P}^2\over \pi M^4}\,.
\end{equation}
One should note that even though the spectral index $n$ in this model
is greater than one, for typical values of the parameters, $n = 1$ with
great accuracy. For example, one may consider the numerical values of
parameters for the version of the hybrid inflation model discussed in
Ref.~\cite{hybrid}: $g^2 \sim \lambda \sim10^{-1}$, $m \sim 10^2$ GeV
(electroweak scale), $M \sim 10^{11}$ GeV. In this case $n-1= {\cal
  O}(10^{-4})$. Thus it is not very easy to use hybrid inflation in
order to obtain a blue spectrum without fine-tuning. One could
even argue that hybrid inflation has a very stable prediction that the
spectrum must be almost exactly flat; see, however, Ref.~\cite{GBW}.

This conclusion would be somewhat premature. Indeed, one can follow
the lines of Ref.~\cite{LinMukh} and consider models where the field
$\phi$ during inflation acquires an effective mass squared
\begin{equation}\label{meff}
m^2_{\rm eff} = m^2 + \alpha H^2 \,,
\end{equation}
with $\alpha = {\cal O}(1)$. This is a very natural assumption which
is true in a large class of supersymmetric models \cite{moduli,Lyth}.
For example, effective potentials of the scalar fields in supergravity
usually grow as an exponent of the square of the scalar field. A
simplest potential of such type (for $\psi = 0$, i.e. during the stage
of inflation in our scenario) is given by
\begin{equation}\label{POT}
V(\phi) = V_0 \,\exp\Big({4\pi\alpha\phi^2\over3M_{\rm P}^2}\Big)\,,
\end{equation}
where $V_0=M^4/4\lambda$ is the vacuum energy, and we are assuming
$4\pi\alpha\phi^2/3 \ll M_{\rm P}^2$. The effective mass of $\phi$
becomes
\begin{equation}\label{MAS}
m^2 = V''(\phi) = \alpha H^2\,\Big(1 +
{8\pi\alpha\phi^2\over3M_{\rm P}^2}\Big) \simeq \alpha H^2\,.
\end{equation}
The field evolves according to Eq.~(\ref{soln}) with $r\simeq
\alpha/3$, and $\dot\phi^2 + m^2\phi^2 \ll V_0$ is satisfied during
inflation.

The same conclusion is also true for the models where the field $\phi$
has a non-minimal coupling with gravity described by the additional
term $-\xi R\phi^2/2$ in the Lagrangian. During hybrid inflation $R =
12 H_0^2$, and the field $\phi$ acquires an effective mass
(\ref{meff}) with $\alpha = 12\xi $. In such models one should
consider those parts of the universe where $\phi \ll M_{\rm
  P}/\sqrt\alpha$, because at much larger values of $\phi$ the
effective gravitational constant $G^{-1} = {M^2_{\rm P}/16\pi -{\xi
    }\phi^2/2}$ becomes singular.  For $\phi \ll M_{\rm P}/
\sqrt\alpha$ the Hubble constant can be approximated by
Eq.~(\ref{H0}).

A generalization of Eq.~(\ref{denspert}) for this case in the regime
$\alpha H^2 \gg m^2$ is
\begin{equation}\label{newden}
\frac{\delta\rho}{\rho} (k) =  {9\sqrt 2 H\over 5\pi \alpha\phi }  =
{3\sqrt{6}\, g M \over 5 \sqrt{\lambda \pi}\, \alpha M_{\rm P}  }
\left({k\over H_0}\right)^{\alpha\over3} \,.
\end{equation}
Here $\phi$ is the value of the scalar field at the moment when the
perturbations are produced which at the end of inflation have momentum
$k$.  These perturbations have spectral index $n - 1 \simeq 2\alpha/3
> 0$ for $\alpha \lesssim 1$. We call such models {\it tilted hybrid
  inflation}: they provide a spectrum of density perturbations with a
positive tilt which may be large, and which may not require any fine
tuning of parameters.

For most applications  it is not very important   which particular
mechanism gives the inflaton mass the contribution proportional to
$\alpha H^2$; one may simply assume that for whatever reason the
effective mass of the inflaton field is given by $m^2_{\rm eff} = m^2
+ \alpha H^2$, and that $ \alpha H^2 \gg m^2$ at the last stages of
inflation. For definiteness, we will concentrate on the model of
Eq.~(\ref{POT}). In this model,
the condition for inflation to occur, $-\dot H \ll H^2$, is
\begin{equation}\label{infl}
\phi < \phi_{\rm inf} \equiv {3\over2\sqrt\pi\,\alpha}\,M_{\rm P}\,.
\end{equation}
As a consequence, the number of $e$-folds, see Eq.~(\ref{soln}),
$N=(3/\alpha)\,\ln(\phi/\phi_c)$, has a maximum value
\begin{equation}\label{Nmax}
N < N_{\rm max} \equiv {3\over\alpha}\,\ln\Big(
{3g\,M_{\rm P}\over2\sqrt\pi\,\alpha M}\Big)\,.
\end{equation}

For this model, the slow-roll parameters are~\cite{LL93}
\begin{eqnarray}
\epsilon &=& {M_{\rm P}^2\over16\pi} \Big({V'\over V}\Big)^2 =
{4\pi\alpha^2\phi^2\over9M_{\rm P}^2} \,, \\[1mm]
\eta &=& {M_{\rm P}^2\over8\pi} \Big({V''\over V}\Big) =
{\alpha\over3}\Big(1+{8\pi\alpha \phi^2\over3M_{\rm P}^2}\Big) \,,
\end{eqnarray}
which gives a spectrum of curvature perturbations
\begin{equation}\label{PR}
{\cal P}_{\cal R}(k) = {4\pi\over M_{\rm P}^2} \Big({H\over2\pi}
\Big)^2 {1\over\epsilon} = {9H^2\over4\pi^2\alpha^2\phi^2}
\end{equation}
where $\phi$ is the value of the scalar field at the moment when those
perturbations which have momentum $k$ at the end of inflation were
produced. This spectrum has a tilt
\begin{equation}\label{tilt}
n-1 = {d\ln{\cal P}_{\cal R}\over d\ln k} =
{2\alpha\over3}\Big(1 -
{4\pi\alpha\phi^2\over3M_{\rm P}^2}\Big)\,.
\end{equation}
Note that the spectrum has a minimum at
\begin{equation}\label{min}
\phi = \phi_{\rm min} \equiv \sqrt{3\over4\pi\alpha}\,M_{\rm P}\,,
\end{equation}
where the tilt is $n=1$ and the spectrum is locally scale invariant.
Later one, it will acquire the asymptotic value $n=1+2\alpha/3$.
This is an interesting feature of our model.

In order to find observational constraints on the parameters of this
model one should consider temperature anisotropies of the microwave
background produced by the scalar (density) and tensor (gravity wave)
perturbations. Quantum fluctuations of the inflaton field $\phi$
produce long-wavelength curvature perturbations. We will use ${\cal
  R}$ to denote the curvature perturbation on comoving hypersurfaces.
In order to compare with observations we have to compute the effect
that such a perturbation has on the temperature anisotropies of the
CMB (expanded in spherical harmonics),
\begin{equation}\label{temp}
{\delta T\over T}(\theta,\phi) = \sum_{lm} a_{lm} \,
Y_{lm}(\theta,\phi)\,.
\end{equation}
The main effect on large scales comes from the ordinary Sachs-Wolfe
effect~\cite{SW67}. The angular power spectrum,
$C_l\equiv\langle|a_{lm}|^2\rangle$, can be computed from the spectrum
of scalar perturbations, ${\cal P}_{\cal R}(k)$, as
\begin{equation}\label{CL}
C_l = 2\pi^2\,\int_0^\infty{dk\over k}\,
{\cal P}_{\cal R}(k)\,I_{kl}^2\,,
\end{equation}
where the `window function' $I_{kl}$ indicates how a given scale $k$
contributes to the $l$-th multipole of the power spectrum and it can
be expressed, for a flat universe, in terms of spherical Bessel
functions~\cite{LL93}. In our model, the spectrum of the scalar
perturbations (evaluated at the end of inflation) is given by
Eq.~(\ref{PR}). For an arbitrary tilt of the spectrum, $n\neq1$, the
power spectrum behaves as
\begin{equation}\label{CL2}
C_l = C_2 \, {\Gamma(l+{n-1\over2})\,\Gamma(4-{n-1\over2})\over
\Gamma(l+2-{n-1\over2})\,\Gamma(2+{n-1\over2})}\,.
\end{equation}
Note that for a nearly scale invariant spectrum, $n\simeq1$, we
recover the familiar result
\begin{equation}\label{CLL}
l(l+1)\,C_l \simeq {2\pi\over25}\,{\cal P}_{\cal R}
\simeq {\rm constant}\,.
\end{equation}

Observations of the first few multipoles of the angular power spectrum
by COBE~\cite{COBE} already constrain $l(l+1)\,C_l \simeq 8\times
10^{-10}$ on large scales. This imposes a constraint on the parameters
of the model,
\begin{equation}\label{scalar}
{3g^2 M^2 \,e^{-2\alpha N/3}\over2\lambda\,\alpha^2\,M_{\rm P}^2}
\simeq 10^{-8}\,,
\end{equation}
where $N=\ln(H_0/k)$ is the number of $e$-folds from the end of
inflation when the scale $k$ crossed outside the horizon, and $\phi_N
= (M/g)\,\exp(\alpha N/3)$ is the value of the scalar field at that
time. The observations made by COBE also constrain the tilt of the
spectrum on large scales, $n=1.2\pm0.3$ at the $2\sigma$
level~\cite{COBE}. In our model, the tilt is $n=1+2\alpha/3$. For
example, for $\alpha = 0.25$ one has $n\simeq1.2$. The number of
$e$-folds associated with scales corresponding to our present horizon
depends only logarithmically on quantities like the reheating
temperature,
\begin{equation}
N \simeq 55 + {2\over3} \ln\Big({V^{1/4}\over10^{16}{\rm GeV}}\Big)
+ {1\over3} \ln\Big({T_{\rm rh}\over10^{9}{\rm GeV}}\Big)\,.
\end{equation}
For $N\simeq 55$, and values of the couplings $g^2 \sim \lambda \sim
0.1$, we have
\begin{equation}\label{bound}
M \simeq 2\times 10^{-3} M_{\rm P} \simeq 2 \times 10^{16} {\rm GeV}\,,
\end{equation}
which is the GUT scale. If the scalar field $\phi$ has a bare mass $m$,
it should be smaller than
\begin{equation}\label{mass}
\sqrt\alpha H_0  \simeq 10^{-5} M_{\rm P}   \,.
\end{equation}
As long as $m$ is smaller than $\sqrt \alpha H_0$, nothing depends on
this parameter, so we do not need to fine-tune $m$ in order to obtain
adiabatic scalar perturbations with a blue spectrum. Since $H_0\ll
M$, hybrid inflation will end with a sudden transition to the symmetry
breaking phase of the triggering field, without a second stage of
inflation like the one discussed in Refs.~\cite{Guth,GBLW}.

Depending on the parameters of the theory and the efficiency of
reheating, the minimum of the spectrum of curvature perturbations
(\ref{min}) for $\alpha \sim 0.25$ could lie on scales comparable to
the scale of our present horizon. If the minimum corresponds to the
horizon scale, the spectrum on this scale will be practically flat.
For the parameters used above, $\alpha=0.25$ and $g^2\simeq \lambda
\simeq 0.1$, we find $N_{\min} \equiv N(\phi_{\rm min}) \sim 60$,
which is similar to the number of $e$-folds corresponding to the scale
of the horizon.  In this case, the scalar spectrum is nearly scale
invariant, $n=1$, see Eq.~(\ref{tilt}), on those scales, but soon
rises to an asymptotic value $n\simeq 1.2$ on smaller scales.  On the
other hand, for $\alpha \gtrsim 0.25$ one may obtain a spectrum with a
minimum at some intermediate scale. This would lead to an increase of
density perturbations on the horizon scale with respect to its
amplitude in the minimum.  However, this could occur only for some
particular choice of the parameters.

Note that in our model inflation is naturally very short. For example,
in the theories with the potential (\ref{POT}) inflation occurs only
for $\phi < \phi_{\rm inf}$, which gives $N < N_{\rm max} \simeq 75$
for $g^2 = 0.1$ and $\alpha = 0.25$, see Eq.~(\ref{Nmax}).  Meanwhile
for $\alpha \sim 0.3$ one has $N_{\rm max} \simeq 60$. It makes this
model very interesting from the point of view of open inflation, where
excessive inflation is undesirable~\cite{GBL}, but simultaneously
implies that $n \sim 1.2 - 1.3$ is about the largest spectral index
that can be obtained in our scenario~\cite{RefMukh}. On the other
hand, observations of large scale structure do not allow a much larger
tilt~\cite{viana}, while constraints from primordial black hole
production~\cite{jim} give a similar bound.

This model also has tensor or gravitational wave perturbations which
could in principle distort the CMB. The amplitude of the tensor
perturbations  spectrum is given by~\cite{LL93,GBW}
\begin{equation}\label{PGW}
{\cal P}_g = {64\pi\over M_{\rm P}^2} \Big({H\over2\pi}\Big)^2
= {16H^2\over\pi M_{\rm P}^2} \,.
\end{equation}
Let us define the ratio of tensor to scalar components of temperature
anisotropies as~\cite{LL93}
\begin{equation}\label{ratio}
R \simeq {3{\cal P}_g\over4{\cal P}_{\cal R}} =
{16\pi\alpha^2\phi^2\over 3M_{\rm P}^2} \,.
\end{equation}
At present there is no observational evidence of any tensor component
in the CMB anisotropies; the observed power spectrum on large scales
is perfectly compatible with a scalar component and no tensors. In the
future, the new generation of satellites will be able to separate the
two components if the ratio is large enough ($R>0.15$), see
Ref.~\cite{COBRAS}. The condition $R\ll1$ together with
Eq.~(\ref{scalar}) gives
\begin{equation}\label{tensor}
{H^2\over M_{\rm P}^2} \ll 8\times10^{-10}\,.
\end{equation}
For $\alpha < 0.2$, $\lambda \sim g^2 \simeq 0.1$ this condition is
satisfied. For greater values of $\alpha$ parameters, one could have a
significant contribution of tensor perturbations to the temperature
anisotropies. For example, for $\phi=\phi_{\rm min}$ at horizon scales
we have $R=4\alpha/3$, and the constraint on the scalar amplitude
(\ref{scalar}) is then more stringent by a factor $(1+R)$.
However, this practically does not change the values of the parameters
of our model for $\alpha = 0.25, \lambda \sim g^2 \sim 0.1$.

In conclusion, we have found a model of hybrid inflation which
provides, in a natural way, a positive tilted spectrum with the
correct amount of scalar and tensor perturbations to be in agreement
with large scale observations of the CMB anisotropies.

The work by A.L. was  supported by  NSF grant PHY-9219345.
This work was also supported by a NATO Collaborative Research Grant,
Ref.~CRG.950760.


\begin{references}

\bibitem{book} A. D. Linde, {\it Particle Physics and Inflationary
    Cosmology} (Harwood, Chur, Switzerland, 1990).

\bibitem{MAP} {\it MAP} Home Page at \\
  {\tt http://map.gsfc.nasa.gov/} (1996).

\bibitem{COBRAS} {\it COBRAS/SAMBA} Home Page at \\
  {\tt http://astro.estec.esa.nl/SA-general/Projects/\\
    Cobras/cobras.html} (1996).

\bibitem{hybrid} A.D. Linde, Phys. Lett.  {\bf B259}, 38 (1991); Phys.
  Rev. D {\bf 49}, 748 (1994).

\bibitem{LL93} A. R. Liddle and D. H. Lyth, Phys. Rep. {\bf 231}, 1
  (1993).

\bibitem{CLLSW} E. J. Copeland, A. R. Liddle, D. H. Lyth, E. D.
  Stewart and D. Wands, Phys. Rev. D {\bf 49}, 6410 (1994).

\bibitem{GBW} J. Garc\'\i a-Bellido and D. Wands, Phys. Rev. D {\bf
    54}, 7181 (1996), {\tt astro-ph/9606047}.

\bibitem{Lyth} For a comprehensive review, see D.H. Lyth, ``Models of
  Inflation and the Spectral Index of the Density Perturbation,'' {\tt
    hep-ph/9609431} (1996).

\bibitem{open} J. R. Gott, Nature, {\bf 295}, 304 (1982); M. Sasaki,
  T. Tanaka, K.  Yamamoto and J. Yokoyama, Phys. Lett. B {\bf 317},
  510 (1993); M.  Bucher, A. S. Goldhaber and N. Turok, Phys. Rev.  D
  {\bf 52}, 3314 (1995).

\bibitem{LM} A. D. Linde, Phys. Lett. {\bf B 351}, 99 (1995); A. D.
  Linde and A. Mezhlumian, Phys. Rev. D {\bf 52}, 6789 (1995).

\bibitem{WS} M. White and J. Silk, {\it Observational Constraints on
  Open Inflation Models}, e-Print Archive: {\tt astro-ph/9608177} (1996).

\bibitem{GBL} J. Garc\'\i a-Bellido and A. D. Linde, in preparation.

\bibitem{Guth} L. Randall, M. Solja\v ci\'c and A. H. Guth, Nucl.
  Phys. {\bf B 472}, 377 (1996).

\bibitem{GBLW} J. Garc\'\i a-Bellido, A. D. Linde and D. Wands, Phys.
  Rev. D {\bf 54}, 6040 (1996), {\tt astro-ph/9605094}.

\bibitem{LinMukh} A. Linde and V. Mukhanov, {\it NonGaussian
    Isocurvature Perturbations from Inflation}, {\tt astro-ph/9610219}
  (1996).

\bibitem{moduli} M. Dine, W. Fischler and D. Nemeschansky, Phys.
  Lett. {\bf B136}, 169 (1984); G.D. Coughlan, R. Holman, P.  Ramond
  and G.G. Ross, Phys. Lett. {\bf B140}, 44 (1984); A.S. Goncharov,
  A.D.  Linde, and M.I. Vysotsky, Phys. Lett.  {\bf 147B}, 279 (1984);
  M. Dine, L.  Randall and S. Thomas, Nucl.  Phys. {\bf 458}, 291
  (1996);

\bibitem{SW67} R. K. Sachs and A. M. Wolfe, Astrophys. J. {\bf 147},
  73 (1967).

\bibitem{COBE} C. L. Bennett, et al., Astrophys. J. {\bf 464}, L1
  (1996).

\bibitem{RefMukh} One can get much greater values of $n$ in the model
  of Ref.~\cite{LinMukh}, both for adiabatic and isothermal perturbations.

\bibitem{viana} A. R. Liddle, D. H. Lyth, R. K. Schaefer, Q. Shafi and
  T. P. Viana, Mon. Not. Roy. Astr. Soc. {\bf 281}, 531 (1996).

\bibitem{jim} B. J. Carr, J. H. Gilbert and J. E. Lidsey, Phys. Rev.
  D {\bf 50}, 4853 (1994).

\end{references}
\end{document}